\documentclass[twocolumn,preprintnumbers,amsmath,amssymb,superscriptaddress,footnote]{revtex4-2}
\usepackage{graphicx} 
\usepackage[dvipdfmx]{epsfig} 
\usepackage{bm}
\usepackage{ulem}
\usepackage{amsmath,color}

\def\m@thcombine#1#2{
  \setbox0=\hbox{$#1$}
  \setbox1=\hbox{$#2$}
  \ifdim\wd0>\wd1
    \setbox0=\hbox to\wd1{\hss\box0\hss}
  \else
    \setbox1=\hbox to\wd0{\hss\box1\hss}
  \fi
  \mathop{\vcenter{
    \offinterlineskip\box0\box1}}}
\def\lesim{\m@thcombine<\sim}
\def\gesim{\m@thcombine>\sim}

\begin{document}
\title{Density profiles near nuclear surface of $^{44,52}$Ti:
An indication of $\alpha$ clustering}
\author{W. Horiuchi}
\email{whoriuchi@omu.ac.jp}
\affiliation{Department of Physics, Osaka Metropolitan University, Osaka 558-8585, Japan}
\affiliation{Nambu Yoichiro Institute of Theoretical and Experimental Physics (NITEP), Osaka Metropolitan University, Osaka 558-8585, Japan}
\affiliation{RIKEN Nishina Center, Wako 351-0198, Japan}
\affiliation{Department of Physics,
  Hokkaido University, Sapporo 060-0810, Japan}

\author{N. Itagaki}
\email{itagaki@omu.ac.jp}
\affiliation{Department of Physics, Osaka Metropolitan University, Osaka 558-8585, Japan}
\affiliation{Nambu Yoichiro Institute of Theoretical and Experimental Physics (NITEP), Osaka Metropolitan University, Osaka 558-8585, Japan}

\preprint{NITEP 147}

\begin{abstract}
  We investigate the degree of $\alpha$ ($^4$He nucleus) clustering
  in the ground-state density profiles of $^{44}$Ti and $^{52}$Ti.
  Two types of density distributions,
  shell- and cluster-model configurations,
  are generated fully microscopically
  with the antisymmetrized quasi-cluster model,
  which can describe both the $j$-$j$ coupling shell
  and $\alpha$-cluster configurations
  in a single scheme.
  Despite both the models reproducing measured charge radius data,
  we found that the $\alpha$ clustering significantly diffuses
  the density profiles near the nuclear surface
  compared to the ideal $j$-$j$ coupling shell model configuration.
  The effect is most significant for $^{44}$Ti, while it is less
  for $^{52}$Ti due to the occupation of the $0f_{7/2}$ orbits
  in the $^{48}$Ca core. This difference
  can be detected by measuring proton-nucleus elastic scattering
  or the total reaction cross section on a carbon target
  at intermediate energies.
\end{abstract}
\maketitle

\section{Introduction}

Nucleon density distributions include various information on
the nuclear structure. Saturation of the nuclear density
in the internal region and the drop off at the nuclear surface
were revealed by systematic
measurements of charge density distributions
using electron scattering~\cite{Hofstadter56}.
The nucleon density distribution can also be obtained
using proton-nucleus elastic scattering~\cite{Sakaguchi17}.
Such measurements were  extended to unstable nuclei
using the inverse kinematics~\cite{Matsuda13}.
Since the nuclear density is saturated in the internal region,
the nuclear structure information is obtained
near the nuclear surface~\cite{Horiuchi20,Horiuchi21}.
For example, nuclear deformation induces
a sudden enhancement of the nuclear matter radius~\cite{Minomo11,Minomo12,Sumi12,Horiuchi12,Takechi12,Takechi14,Watanabe14,Horiuchi15,Horiuchi22},
where the density profile near the nuclear surface is
significantly diffused compared to
the spherical configuration~\cite{Hatakeyama18,Choudhary21}.
The nuclear ``bubble'' structure, the internal density depression
can also be imprinted on the nuclear surface~\cite{Choudhary20}.

We explore how the nuclear structure affects the density profiles
near the nuclear surface.
These days, exploring an $\alpha$ ($^{4}$He nucleus) cluster
in medium to heavy mass nuclei has attracted attention
in the context of the astrophysical interest~\cite{Typel10}.
Direct measurement of the degree of $\alpha$ clustering
near the nuclear surface has been realized
using $\alpha$ knockout reactions~\cite{Tanaka21}.
The quantification of the degree of $\alpha$ clustering
may impact the determination of the reaction rates
of astrophysically important reactions involving medium mass nuclei.

Although the clustering is exotic and intriguing to explore, the standard picture for the nuclear structure is shell structure,
and the difference between these two
must appear in the density profiles near the nuclear surface.
Here we choose $^{44}$Ti and $^{52}$Ti as representatives of medium mass nuclei.
The well-developed $^{40}{\rm Ca}+\alpha$
structure of $^{44}$Ti was predicted in Ref.~\cite{Michel86}.
Afterward, the inversion doublet structure
was confirmed experimentally~\cite{Yamaya90,Yamaya98}
as its supporting evidence.
The $\alpha$-cluster structure of $^{44}$Ti was microscopically
investigated~\cite{Kimura06}.
Establishing the degree of the clustering in $^{44}$Ti
may impact $^{40}{\rm Ca}(\alpha,\gamma)^{44}{\rm Ti}$ reaction rate~\cite{Nassar06}.
The influence of the $\alpha$ clustering
on the reaction rate was discussed for $^{48}$Ti using the $(p,p\alpha)$
knockout
  reactions~\cite{Taniguchi21}.
We remark that the mechanism of the emergence of
the $\alpha$ cluster near the nuclear surface in medium mass nuclei is
recently suggested concerning the tensor force~\cite{Ishizuka22}.

In this paper, we discuss the difference of the density
profiles near the nuclear surface  between cluster
and shell models
by taking an example of $^{44}$Ti. We also examine the case of $^{52}$Ti
to clarify the role of excess neutrons.
The study along this line may give a hint for the research for
the emergent mechanism
of $\alpha$ particle in the neutron-rich nuclei
toward understanding nuclear matter properties.
For this purpose, we need a model that can describe
both the shell and cluster configurations in a single scheme.
Here we employ the antisymmetrized quasi-cluster model (AQCM~\cite{AQCM01,AQCM02,AQCM03,AQCM04,AQCM05,AQCM06,AQCM07,AQCM08,AQCM09,AQCM10,AQCM11,AQCM12,AQCM13,AQCM14}). This model allows to smoothly transform the cluster model wave function to the $j$-$j$ coupling shell one and these two can be treated
on the same footing.

The paper is organized as follows.
Section~\ref{methods.sec} summarizes the present approach
to investigate the $\alpha$ clustering in the density profiles
of $^{44}$Ti and $^{52}$Ti.
How to calculate the density distributions
that have shell and cluster configurations using
the AQCM is explained in Sec.~\ref{AQCM.sec}.
For the sake of convenience,
some definitions of the nuclear radii are given in Sec.~\ref{radii.sec}.
To connect obtained density profiles with reaction observables,
a high-energy reaction theory,
the Glauber model is briefly explained in Sec.~\ref{Glauber.sec}.
Section~\ref{results.sec} presents our results.
First, in Sec.~\ref{wf.sec}, we discuss the properties of the wave functions
with the shell and cluster configurations.
Definitions and characteristics of
the two types of model wave functions are described in detail.
In Sec.~\ref{density.sec}, we compare the resulting density profiles
and discuss the relationship between these density profiles
and reaction observables.
Section~\ref{density2.sec} clarifies the difference in
the shell- and cluster-model approaches in the density profiles.
Finally, the conclusion is given in Sec.~\ref{conclusion.sec}.

\section{Methods}
\label{methods.sec}

\subsection{Density distribution with antisymmetrized quasi-cluster model (AQCM)}
\label{AQCM.sec}

The AQCM ansatz of the core ($^{40}$Ca or $^{48}$Ca)
plus $\alpha$ particle
wave function, which can be transformed to $j$-$j$ coupling shell model one,
is defined by
fully antisymmetric $(\mathcal{A})$ product
of the core and $\alpha$ wave functions as
\begin{align}
  \Phi(\nu_{\rm C},\nu_\alpha,R,\Lambda_p,\Lambda_n)=\mathcal{A}\left\{\Phi_{\rm C}(\nu_{\rm C})\Phi_{\alpha}(\nu_\alpha,R,\Lambda_p,\Lambda_n)\right\}.
\end{align}
The wave function of the core nucleus $\Phi_C$ with
the oscillator size parameter $\nu_C$ is constructed
based on the multi-$\alpha$ cluster model~\cite{Brink}.
For $^{40}$Ca, the core wave function is
obtained by taking small distances among ten $\alpha$ clusters;
this nucleus corresponds to the closure of the $sd$-shell and
the shell- and cluster-model wave functions coincide at the zero-distance
limit of the inter-cluster distances. 
 For $^{48}$Ca, we need to put additional eight neutrons
 describing the neutron number $N=28$, subclosure of the $0f_{7/2}$ shell, 
and AQCM allows a simple description to transform the cluster model.
The details are given in Ref.~\cite{AQCM06}.

The wave function of the $\alpha$ particle at the distance
between the center-of-mass coordinate of the core and $\alpha$ particles,
$R$, is defined as
the product of the single-particle Gaussian wave packet as
\begin{align}
  \Phi_{\alpha}(\nu,R,\Lambda_p,\Lambda_n)=
  \phi^\nu_1(\uparrow,p)\phi^\nu_2(\downarrow,p)\phi^\nu_3(\uparrow,n)\phi^\nu_4(\downarrow,n)
\end{align}
with a single-nucleon Gaussian wave packet
with spin $\chi_s$ ($s=\uparrow$ or $\downarrow$)
and isospin $\eta_t$ ($t=p$ or $n$) wave functions
\begin{align}
  \phi^\nu_i(s,t)=  \left(\frac{2\nu}{\pi}\right)^{3/4}\exp\left[-\nu(\bm{r}_i-\bm{\zeta}_{t})^2\right]\chi_{s}\eta_{t},
\end{align}
where
\begin{align}
  \bm{\zeta}_t=\bm{R}+i\Lambda_t \bm{e}_t^{\rm spin}\times \bm{R}
\end{align}
with $\bm{e}^{\rm spin}_t$ being a unit vector for the intrinsic-spin
orientation of a nucleon. 
Note that it corresponds to the ordinary Brink $\alpha$-cluster wave function 
in Ref.~\cite{Brink} by taking $\Lambda_t=0$.
A limit of $R\to 0$ leads to the SU(3) limit of the shell model configuration.
The $j$-$j$ coupling shell model wave function can be expressed
by introducing $\Lambda_t=1$ with $R\to 0$~\cite{AQCM04}.
For example, in $^{44}$Ti, the $\alpha$ cluster
  is changed into $(0f_{7/2})^4$ configuration using AQCM. 
Thus, the model wave function can describe both the shell and
$\alpha$-cluster configurations in a single scheme.

Finally, the density distribution in the laboratory frame
is obtained by averaging the intrinsic density distribution over angles as
\begin{align}
  \rho_t(r)=\frac{1}{4\pi}\,\int d\hat{\bm{r}}\,\rho^{\rm int}_{t}(\bm{r}),
\end{align}
where $\rho^{\rm int}_t$ is obtained by using the Slater determinant of $^{44}$Ti or $^{52}$Ti represented as $\Phi$
\begin{align}
  \rho^{\rm int}_{t}(\bm{r}) = \langle \Phi |\sum_{i\in t} \delta(\bm{r}_i-\bm{r}) | \Phi \rangle / \langle \Phi | \Phi \rangle,
\end{align}
where the summation is taken over protons ($t=p$) or neutrons ($t=n$).
Note that $\sum_{i=1}^A\left<\bm{r}_i\right>=0$ is imposed and
the center-of-mass motion is ignored as the mass number
$A\approx 40$--50 is large.

\subsection{Definitions of radii}
\label{radii.sec}

The root-mean-square (rms) point-proton, neutron, and matter radii
are calculated by
\begin{align}
  r_p&=\sqrt{\frac{4\pi}{Z}\int_0^\infty dr\, r^4 \rho_p(r)}, \\
  r_n&=\sqrt{\frac{4\pi}{N}\int_0^\infty dr\, r^4 \rho_n(r)},\\
    r_m&=\sqrt{\frac{4\pi}{A}\int_0^\infty dr\, r^4 [\rho_p(r)+\rho_n(r)]},
\end{align}
where $Z$ denotes the proton number.
The charge radius $r_{\rm ch}$ is converted
from the theoretical point-proton radius $r_p$ by
 using the formula~\cite{Friar97,Angeli13}
\begin{align}
r_{\rm ch}^2=r_{p}^2+r_{{\rm ch},p}^2+\frac{N}{Z}r^2_{{\rm ch},n}+
  \frac{3\hbar^2}{4m_p^2c^2}.
\label{charge.eq}  
\end{align}
where $r^2_{{\rm ch},t}$ is the second moment of the nucleon charge
distribution, and the fourth term of Eq.~(\ref{charge.eq}) is
the so-called Darwin-Foldy term,
which comes from relativistic correction.

\subsection{Reaction observables within the Glauber model}
\label{Glauber.sec}

Proton-nucleus elastic scattering at intermediate energy
is one of the most direct ways to extract the density profiles
near the nuclear surface. We remark that 
the whole density distribution can be obtained
by measuring up to backward angles~\cite{Terashima08,Zenihiro10},
although the internal density has large uncertainties.
As long as the nuclear surface density is of interest,
only the cross sections at the forward angles,
to be more specific, the cross section at the first peak in proton-nucleus
diffraction is needed to extract the ``diffuseness'' of the
density distribution as prescribed in Ref.~\cite{Hatakeyama18}.
To connect the density profile with observables
at intermediate energies,
we employ a high-energy microscopic reaction theory,
the Glauber model~\cite{Glauber}.

The elastic scattering differential cross section is evaluated by
\begin{align}
  \frac{d\sigma}{d\Omega}=|f(\theta)|^2
\end{align}
with the scattering amplitude of the proton-nucleus elastic scattering~\cite{Suzuki03}
\begin{align}
  f(\theta)=F_C(\theta)+\frac{ik}{2\pi}\int d\bm{b}\, e^{-i\bm{q}\cdot\bm{b}+2i\eta \ln(kb)}\left(1-e^{i\chi_{pT}(\bm{b})}\right),
\end{align}
where $F_C(\theta)$ is the Rutherford scattering amplitude,
$\bm{b}$ is the impact parameter vector,
and $\eta$ is the Sommerfeld parameter.
The relativistic kinematics is used for the wave number $k$.

The optical phase-shift function $\chi_{pT}$ includes all dynamical information
in the Glauber model, but its evaluation involves
multi-fold integrations. For practical calculations,
the optical-limit approximation (OLA)~\cite{Glauber,Suzuki03}
is made to compute
the optical phase-shift function as
\begin{align}
  i\chi_{pT}(\bm{b})\approx -\int d\bm{r}\,
  \left[\rho_p(\bm{r})\Gamma_{pp}(\bm{b}+\bm{s})
  +\rho_n(\bm{r})\Gamma_{np}(\bm{b}+\bm{s})\right],
\end{align}
where $\bm{r}=(\bm{s},z)$ with $z$ being the beam direction.
The inputs to the theory are the projectile's density
distributions and proton-proton (neutron-proton) profile function
$\Gamma_{pp}$ ($\Gamma_{np}$).
The parameterization of the profile function is given
in Ref.~\cite{Ibrahim08}.
Once all the inputs are set, the theory has no adjustable parameter,
and thus, the resulting reaction observables must reflect
the density profiles of the projectile nucleus.
The OLA works well for proton-nucleus scattering
as demonstrated, e.g., in Refs.~\cite{Horiuchi16, Hatakeyama19},
and its accuracy compared to those obtained by the full evaluation
of the optical phase-shift function
were discussed in Refs.~\cite{Varga02,Ibrahim09,Nagahisa18,Hatakeyama19}.

  The density profile can also be reflected in
  the total reaction cross sections at medium to high incident energies,
  which are a standard physical quantity
  to investigate the nuclear size properties.
  Here, we investigate the total reaction cross sections
  on a carbon target as a carbon target is superior
  than a proton target to probe the density distributions
  near the nuclear surface~\cite{Horiuchi14,Makiguchi22}.
In the Glauber model~\cite{Glauber},
the cross section is calculated as
\begin{align}
  \sigma_R=\int d\bm{b}\, \left(1-|e^{i\chi_{PT}(\bm{b})}|^2\right).
\end{align}
Since the multiple scattering effects cannot be neglected in
the nucleus-nucleus collision,
the nucleon-target formalism in the Glauber model~\cite{NTG}
is employed to evaluate projectile-target optical phase-shift
function $\chi_{PT}(\bm{b})$.
The inputs to the theory are the density distributions
of the projectile and target and the profile function.
We take harmonic-oscillator type density for the target density
that reproduces the measured charge radius of $^{12}$C~\cite{Angeli13}.
This model works well in many examples of the nucleus-nucleus
scattering involving unstable nuclei~\cite{Horiuchi06,Horiuchi07,Ibrahim09,Horiuchi10,Horiuchi12,Horiuchi15,Nagahisa18}
and is a standard tool to extract nuclear size properties
from the interaction cross section
measurements~\cite{Kanungo10,Kanungo11,Bagchi20}.

\section{Results and discussions}
\label{results.sec}

\subsection{Properties of the wave functions}
\label{wf.sec}

Here we examine two types of model wave functions:
one is a shell-model-like configuration (S-type),
and another is a cluster-model-like configuration (C-type).
Both models reproduce the experimental charge radius data of $^{44}$Ti.
To clarify the role of the excess neutrons, we examine $^{52}$Ti as well.
Note that the charge radius of $^{52}$Ti has not been measured yet,
and thus we use the data of the neighboring nucleus, $^{50}$Ti,
3.57 fm \cite{Angeli13} as a reference.
In the following two subsubsections,
we explain how to construct the two model wave functions in detail.

\subsubsection{Shell-model-like configuration (S-type)}

The shell-model-like wave function (S-type)
is practically constructed by taking the core-$\alpha$ distance $R$ small
with $\nu_C=\nu_\alpha=\nu$.
It is known that this limit goes to the SU(3) shell model configuration~\cite{Brink}.
For $^{44}$Ti, to
 express the $j$-$j$ coupling shell model wave function,
we take $\Lambda_p=\Lambda_n=1$~\cite{Ishizuka22}, and thus
the wave function of the valence nucleon orbit
becomes $(0f_{7/2})^2_p(0f_{7/2})^2_n$,
where $p$ is for proton and $n$ is for neutron.
In this S-type wave function, as we fix the core-$\alpha$
distance small, we only have one parameter, 
the oscillator size parameter of $^{44}$Ti, $\nu$.
This is fixed to reproduce the point-proton radius
extracted from the charge radius data of $^{44}$Ti.
To confirm the configurations are all right,
we evaluate the total harmonic oscillator quanta $\left<Q\right>$,
the expectation values of single-particle spin-orbit operators
$\sum_{i=1}^A\bm{l}_i\cdot\bm{s}_i$, $\left<LS\right>$,
and single-particle parity operators
$\sum_{i=1}^AP_{i}$ with $P_if(\bm{r}_i)=f(-\bm{r}_i)$,
$\left<P\right>$. The last quantity represents difference of number of particles in the positive-parity orbits and negative-parity orbits.
These calculated values are listed in Table~\ref{results.tab}
and perfectly agree with the results expected
from ideal shell model configurations:
$\left<Q\right>=60$, $\left<LS\right>=0$, and $\left<P\right>=16$
for $^{40}$Ca with the closed $sd$ shell
and 72, 6, and 12 for $^{44}$Ti with
the $(0f_{7/2})^2_p(0f_{7/2})^2_n$ configuration.

For $^{48}$Ca, these values also agree with the ideal values
of the shell model,
$\left<Q\right>=84$, $\left<LS\right>=12$, and $\left<P\right>=8$.
In the case  of $^{52}$Ti, as it differs from the case of $^{44}$Ti,
we take $\Lambda_p=1$ and $\Lambda_n=0.5$,
resulting in the desired expectation values
$\left<Q\right>=96$, $\left<LS\right>=16$, and $\left<P\right>=4$
for the $(0f_{7/2})^2_p(1p_{3/2})^2_n$ configuration.
This is because the $0f_{7/2}$ neutron orbit is already filled by
the core nucleus. The additional two neutrons
are found to occupy higher $j$-upper orbits such as $0g_{9/2}$
when $\Lambda_n=1$. In fact, we get $\left<Q\right>\approx 98$
when $\Lambda_p=\Lambda_n=1$ is taken.
As the charge radius of $^{52}$Ti is unknown,
we also generate the $^{52}$Ti wave function by
extending the point-proton radius $r_p$ by 0.05 fm,
which is listed as ``extended'' S-type.

\subsubsection{$\alpha$-cluster-like configuration (C-type)}

\begin{table*}[htb]
  \begin{center}
    \caption{Properties of the shell-model-like (S-type)
      and $\alpha$-cluster-model-like (C-type) wave functions. Values in parentheses are obtained with ideal configurations.}
\begin{tabular}{lccccccccc}
\hline\hline
  &&$\nu$ (fm$^{-2}$) &$R$ (fm)&$\Lambda_p$&$\Lambda_n$ &$\left<Q\right>$ &$\left<LS\right>$&$\left<P\right>$&$r_{\rm ch}$ (fm)\\
\hline
$^{40}$Ca&&0.1315&--&--&--&60.0 (60) &0.0 (0)&16.0 (16) & 3.478\\
$^{44}$Ti (S-type)&&0.1270&0.20&1&1&72.0 (72)&6.0 (6)&12.0 (12) & 3.611\\
$^{44}$Ti (C-type)&&0.1315&2.85&0&0&72.9 & 0.0 (0) &13.2 &3.611\\
\hline
$^{48}$Ca&&0.1311&--&--&--&84.2 (84) &12.0 (12)&8.4 (8)&3.476\\
$^{52}$Ti (S-type)&&0.1297&0.20&1&0.5&96.2 (96)&15.5 (16)& 4.4 (4) &3.569\\
$^{52}$Ti (C-type)&&0.1311&1.60&0&0&96.5&11.2&4.9&3.569\\
$^{52}$Ti (extended, S-type)&&0.1260&0.20&1&0.5&96.2 (96)&15.5 (16)& 4.4 (4) &3.619\\
$^{52}$Ti (extended, C-type)&&0.1311&3.01&0&0&97.4&11.4&5.9&3.619\\
\hline\hline
\end{tabular}  
\label{results.tab}
\end{center}
\end{table*}

The $\alpha$-cluster-model wave function (C-type) is constructed
based on the core plus $\alpha$-cluster model.
In this case, we take $\Lambda_p=\Lambda_n=0$, where
the four nucleons are localized at a distance $R$ from the core nucleus.
The size parameters of the core wave functions 
are respectively fixed to reproduce the charge radii of $^{40,48}$Ca.
For the C-type wave functions of  $^{44,52}$Ti, for
the sake of simplicity, we set $\nu_{\rm C}=\nu_\alpha=\nu$.
This is reasonable because $\alpha$-particle near
the nuclear surface can be distorted by
the interaction and Pauli principle from the core.
In fact, the size of $\alpha$-particle is somewhat enlarged
compared to that in vacuum~\cite{Horiuchi14b}.
Finally, the distances between the core and $\alpha$ particle, $R$,
of $^{44}$Ti and $^{52}$Ti are respectively
fixed to reproduce their charge radii.
Hereafter we refer to this model as C-type.

Table~\ref{results.tab} also lists the properties of the C-type
wave functions.
For $^{44}$Ti, the $\left<LS\right>$ values are zero
as $\Lambda_p=\Lambda_n=0$.
The core and cluster distance is determined to be $R=2.85$ fm,
implying well-developed $\alpha$ clustering near the nuclear surface.
$\left<Q\right>$ is a bit larger than that of the S-type.
This is due to the mixing of $sdg$-shell orbit
($\left<Q_i\right>=4, \left<P_i\right>=+1$),
which can be confirmed from the fact that
the $\left<P\right>$ value of the C-type is larger than that of the S-type.

For $^{52}$Ti, the core-$\alpha$ distance is found to be smaller,
$R=1.60$ fm
to reproduce the charge radius of $^{50}$Ti.
The S-type and C-type wave functions give
similar $\left<Q\right>$ values, while $\left<LS\right>$ value
is reduced
for the C-type wave function because
$\alpha$ cluster part does not contribute to this value.

The distance becomes comparable to that of $^{44}$Ti, $R\approx 3$ fm
when the extended charge radius is assumed for $^{52}$Ti.
In that case, an increase of the $\left<Q\right>$ and $\left<P\right>$
values are attained by the contribution of higher shell,
which is the same reason found in $^{44}$Ti.
The $\left<LS\right>$ value is also reduced for the C-type
compared to the S-type by the same amount as in the case
of those reproducing the charge radius of $^{50}$Ti
because the S-type wave function includes $1p_{3/2}$ orbits, while
the C-type wave function has 
 no contribution from the $\alpha$ cluster part.

\subsection{Density profiles and reaction observables}
\label{density.sec}

\begin{table}[htb]
  \begin{center}
    \caption{Rms point-proton, neutron, and matter
      radii and diffuseness parameters for proton,
      neutron, and matter in units of fm.}
\begin{tabular}{lccccccc}
\hline\hline
  &&$r_p$&$r_n$&$r_m$&$a_p$&$a_n$&$a_m$\\
\hline
$^{40}$Ca&&3.38&3.38&3.38&0.551&0.551&0.551\\
$^{44}$Ti (S-type)&&3.51&3.51&3.51&0.557&0.557&0.557\\
$^{44}$Ti (C-type)&&3.51&3.51&3.51&0.625&0.625&0.625\\
\hline
$^{48}$Ca         &&3.38&3.62&3.52&0.552&0.528&0.540\\
$^{52}$Ti (S-type)&&3.48&3.68&3.59&0.552&0.574&0.572\\
$^{52}$Ti (C-type)&&3.48&3.67&3.59&0.608&0.566&0.593\\
$^{52}$Ti (extended, S-type)&&3.53&3.73&3.65&0.558&0.584&0.579\\
$^{52}$Ti (extended, C-type)&&3.53&3.71&3.63&0.630&0.579&0.606\\
\hline\hline
\end{tabular}  
\label{results2.tab}
\end{center}
\end{table}

Here we investigate the difference between these density profiles obtained
in the previous section.
Table~\ref{results2.tab} lists the root-mean-square (rms)
point-proton $(r_p)$, neutron $(r_n)$, and matter ($r_m$)
radii of these density models employed in this paper.
Thus far, we have obtained different density profiles that have
the same charge radius, i.e., the rms point-proton radius.
For $^{40}$Ca and $^{44}$Ti, as the number of protons and neutrons
are the same, the rms point-neutron radius is the same as that for the protons
by the definition of the AQCM ansatz.
For $^{52}$Ti, the $r_n$ value of the C-type 
are slightly smaller than that of the S-type,
since an $\alpha$-particle is isoscalar and has no neutron-skin thickness
in this model wave function.
This is consistent with the results showing
the negative correlations between the neutron skin-thickness and
$\alpha$-clustering~\cite{Typel10,Zhao21}.

\begin{figure}[ht]
\begin{center}
  \epsfig{file=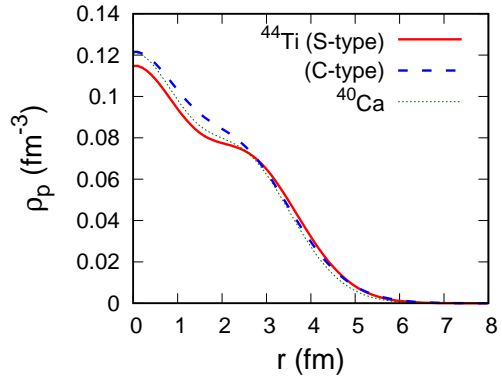, scale=1.2}
  \caption{Point-proton
    density distributions of $^{44}$Ti and $^{40}$Ca.
  The distributions are the same for neutron.}
    \label{dens44Ti.fig}
  \end{center}
\end{figure}

Figure~\ref{dens44Ti.fig} displays the point-proton density distributions
(S-type and C-type)
of $^{44}$Ti. Note that the distributions are the same for the neutrons.
The density distribution of $^{40}$Ca is also plotted for comparison.
Despite that the S-type and C-type density distributions give the same charge radii,
they exhibit different density profiles.
All three densities coincide at $r\approx 3$ fm, which divides
the internal and outer parts of the density distribution.
The internal densities are reduced in the S-type.
This is attributed to the fact that in S-type,
the increase of the charge radius from $^{40}$Ca to $^{44}$Ti
partially comes from the change of
the size of the oscillator parameter, $1/\sqrt{\nu}$.
This leads to the depression of the internal density.
For the C-type, the internal density at around
$r \approx 1$--3 fm is enhanced, which is reasonable,
given the $\alpha$ cluster is located at $r \approx 3$ fm.
Around the surface regions,
at $r \gtrsim 3$~fm the S-type density has larger values, but the inversion occurs, and 
the C-type density is larger
at $r \gtrsim 4$~fm.

\begin{figure}[ht]
\begin{center}
  \epsfig{file=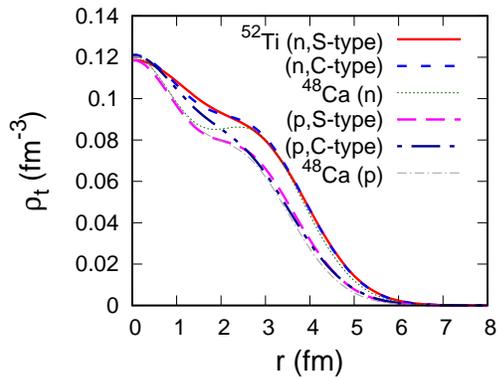, scale=1.2}
  \caption{Point-proton ($t=p$) and neutron ($t=n$) density distributions
    of $^{52}$Ti and $^{48}$Ca.}
    \label{dens52Ti.fig}
  \end{center}
\end{figure}

Figure~\ref{dens52Ti.fig} plots
the point-proton and neutron density distributions
of $^{52}$Ti and $^{48}$Ca.
The charge radius of $^{50}$Ti is used for $^{52}$Ti
to determine the parameters.
While changes in the proton density distributions
from $^{48}$Ca to $^{52}$Ti are small,
they are similar to these for the $^{44}$Ti case.
For the neutron density distributions,
though the densities at the surface region of $r\gtrsim 4$ fm is a little
enhanced, the S- and C-type distributions are quite similar,
implying the effect of the $(0f_{7/2})_n^8$ configuration in
the $^{48}$Ca part. We will address this reason in the next subsection.

\begin{figure}[ht]
\begin{center}
  \epsfig{file=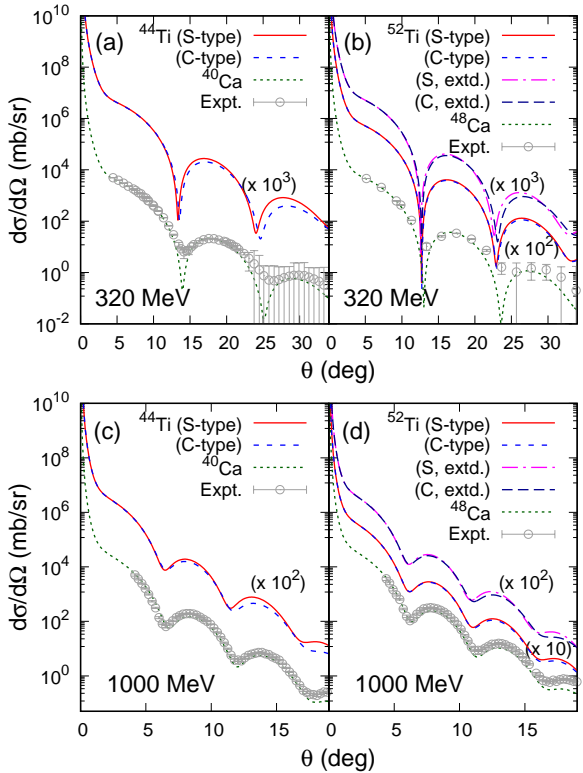, scale=1}                    
  \caption{Differential elastic scattering cross sections
    of (a,c) $^{44}{\rm Ti}+p$ and $^{40}{\rm Ca}+p$
    and (b,d) $^{52}{\rm Ti}+p$ and $^{48}{\rm Ca}+p$ at incident energies of
    (a,b) 320 MeV and (c,d) 1000 MeV as a function of scattering angles.
    The experimental data is taken from Refs.~\cite{Alkhazov76,Kelly91,Feldman94}. For the sake of visibility, the cross sections
      of $^{44}{\rm Ti}+p$ and $^{52}{\rm Ti}+p$ are multiplied by
    some factors. }
    \label{dcs.fig}
  \end{center}
\end{figure}

\begin{figure}[ht]
\begin{center}
  \epsfig{file=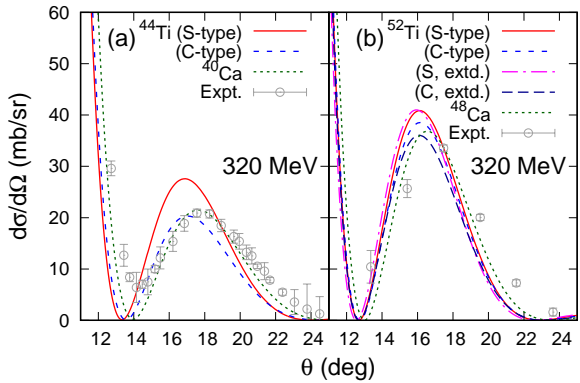, scale=1}                    
  \caption{Same as Fig.~\ref{dcs.fig} (a) and (b) but in a linear scale.}
    \label{dcslinear.fig}
  \end{center}
\end{figure}

These differences between the S-type and C-type are reflected in patterns of proton-nucleus diffraction.
Figure~\ref{dcs.fig} plots the proton-nucleus
differential elastic scattering cross sections.
The proton incident energies are chosen as at 320 and 1000 MeV,
where the experimental data of $^{40,48}{\rm Ca}+p$
is available~\cite{Alkhazov76,Kelly91,Feldman94} (crosses).
Our results perfectly reproduce the data of $^{40,48}{\rm Ca}+p$
 up to the second peak, which verifies our approach.
For $^{44}$Ti, the difference between the two types of density models
(S-type and C-type) is apparent at the first and second peak positions.
For a closer comparison,
  we plot in Fig.~\ref{dcslinear.fig}
  the cross sections in a linear scale.
  We clearly see that the difference between the cross sections
  of the S-type and C-type density models is larger than the uncertainties of
  the experimental $^{40}{\rm Ca}+p$ cross sections at the first peak position.
Measurement of these cross sections is useful
to distinguish the degree of the clustering near the nuclear surface.
In contrast, less difference is found in the cases of $^{52}$Ti
as expected from Fig.~\ref{dens52Ti.fig}.
The difference is found to be comparable to the uncertainties
of the experimental $^{48}{\rm Ca}+p$ cross sections.
The situation is improved when we take the extended charge radius for $^{52}$Ti.

\begin{figure}[ht]
\begin{center}
  \epsfig{file=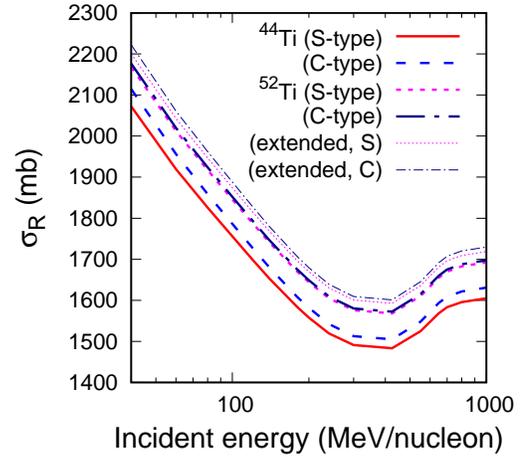, scale=1.2}                    
  \caption{Total reaction cross sections of $^{44}$Ti and $^{52}$Ti
    on a carbon target as a function of incident energy.}
    \label{rcs.fig}
  \end{center}
\end{figure}

These differences in the density profiles
can also influence the total reaction cross sections.
Figure~\ref{rcs.fig} displays the calculated
total reaction cross sections as a function of the incident energy.
Though the difference is not as significant as that in the proton-nucleus
  differential elastic scattering cross sections,
  the difference between the two density models (S-type and C-type)
is at most about 2\% for $^{44}$Ti, which is larger than
the present experimental precision,
typically less than 1\%~\cite{Bagchi20,Tanaka20}.
The cross sections with the S- and C-type density distributions
are almost identical for $^{52}$Ti. A little difference is found when
the extended charge radius is applied to $^{52}$Ti.

To explore the $\alpha$-clustering for heavier nuclei,
we investigate cases for Sn isotopes; to be more specific,
$^{120}{\rm Sn}+\alpha$ and $^{132}{\rm Sn}+\alpha$.
We found that The C-type configuration always
gives more diffused nuclear surface than that of S-type
as we have shown in Ti isotopes.
However, the density profiles of the S- and C-types
becomes similar as the mass number increases
that cannot be distinguished clearly. In such a case, a more direct way, e.g.,
$\alpha$-knockout reaction~\cite{Tanaka21}
could be more useful to quantify the degree of the $\alpha$ clustering.

\subsection{Close comparison of the density profiles}
\label{density2.sec}

To clarify the origin of the differences in the density profiles,
it is convenient to quantify the density profiles near the nuclear surface.
For this purpose, we extract the nuclear diffuseness from the calculated
density distributions using the prescription given in Ref.~\cite{Hatakeyama18}.
Nuclear diffuseness is defined
in a two-parameter Fermi (2pF) function
\begin{align}
\rho_{\rm 2pF}(\bar{R}_q,a_q,r)=\frac{\rho_{0q}}{1+\exp[(r-\bar{R}_q)/a_q]},
\end{align}
where the radius $\bar{R}_q$ and diffuseness $a_q$ parameters
are respectively defined for neutron $(q=n)$, proton $(q=p)$,
and matter ($q=m$). Given the $\bar{R}_q$ and $a_q$ values,
the $\rho_{0q}$ value is uniquely determined
by the normalization condition.
These parameters are determined by minimizing
\begin{align}
  \int_{0}^\infty dr\,r^2\left|\rho_{\rm 2pF}(\bar{R}_q,a_q,r)-\rho_{q}(r)\right|.
\end{align}
Note $\rho_m=\rho_p+\rho_n$.
The extracted 2pF parameters are equivalent to
those obtained by fitting the first peak position and its magnitude
of proton-nucleus elastic
scattering~\cite{Hatakeyama18,Choudhary20,Choudhary21}.

Table~\ref{results2.tab} also lists
the extracted diffuseness parameters for proton, neutron, and matter
density distributions.
These values capture well the characteristics
of the density distributions.
The diffuseness parameters
are similar for $^{40}$Ca and $^{44}$Ti (S-type), while
it is significantly enhanced for $^{44}$Ti (C-type).
In the case of $^{52}$Ti, both the S- and C-types
shows enhanced diffuseness parameters for neutron and matter,
while for proton, a similar behavior is found as that for $^{44}$Ti.

\begin{figure}[ht]
\begin{center}
  \epsfig{file=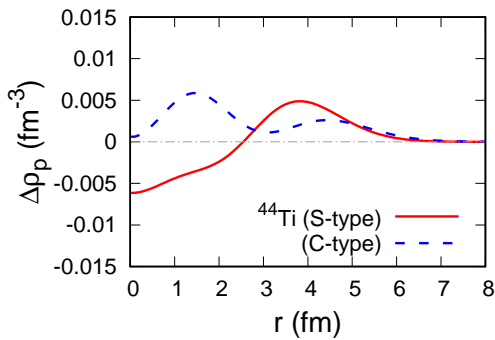, scale=1.2}
  \caption{Difference of point-proton density distributions
  between   $^{44}$Ti and $^{40}$Ca.}
    \label{dens44Tidiff.fig}
  \end{center}
\end{figure}

To verify the reason, we plot in Fig.~\ref{dens44Tidiff.fig}
the difference of the proton density distributions between
$^{44}$Ti and $^{40}$Ca as a function of $r$, i.e.,
$\Delta \rho_p(r)=\rho_p(^{44}{\rm Ti},r)-\rho_p(^{40}{\rm Ca},r)$.
In the S-type, the internal density is depressed 
reflecting the difference of
the oscillator parameters of these nuclei.
It peaks at $r\approx 4$ fm coming from
the additional $(0f_{7/2})_p^2$ configuration.
The $\Delta \rho_p(r)$ value of the C-type behaves quite differently,
showing two peak structure that indicates the inclusion of
the nodal $1p$ orbits, which significantly enhances
the diffuseness of the nuclear surface~\cite{Horiuchi21b}.
The four valence nucleons mainly occupy a ``sharp'' $0f_{7/2}$ orbit
in the S-type density, while a ``diffused'' $1p$ orbit is filled
in the C-type density, leading to the significant difference
in the density profiles near the nuclear surface.

\begin{figure}[ht]
\begin{center}
      \epsfig{file=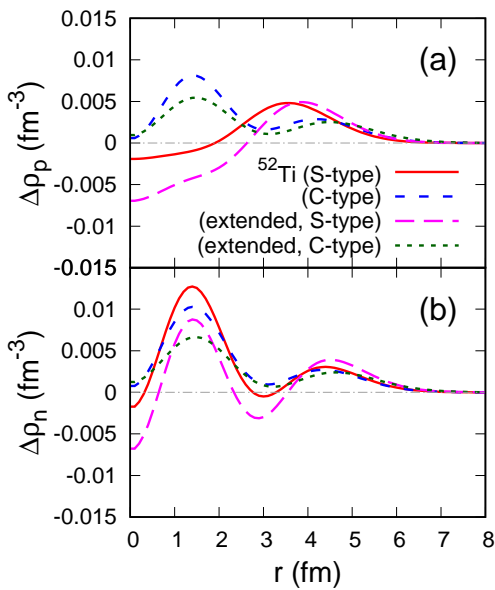, scale=1.2}                    
    \caption{Difference of the density distributions
for (a) proton and (b) neutron between $^{52}$Ti from $^{48}$Ca.}
    \label{dens52Tidiff.fig}
  \end{center}
\end{figure}

For $^{48}$Ca, the diffuseness parameter is smaller than that of $^{40}$Ca
as the sharp $0f_{7/2}$ neutron orbit is filled
as seen in Table~\ref{results2.tab}.
Differently from the $^{44}$Ti case, in $^{52}$Ti,
the neutron diffuseness is enhanced also for the S-type
because the two valence neutrons are considered to occupy
the $1p_{3/2}$ orbit;
$0f_{7/2}$ orbits for the neutrons are fully occupied in the $^{48}$Ca core.
This effect leads to the enhancement of nuclear diffuseness.
Figure~\ref{dens52Tidiff.fig} plots the differences of
the proton and neutron density distributions between $^{52}$Ti and $^{48}$Ca.
The C-type density of $^{52}$Ti behaves like $^{44}$Ti
but the amplitudes in the internal region are larger because the resulting
core-$\alpha$ distance is smaller than that of $^{44}$Ti as we see in Table~\ref{results.tab}. We also calculate the $\Delta \rho_p$ and $\Delta \rho_n$
for the extended $^{52}$Ti density distributions.
The enhancement of the surface region is more apparent
relative to that of the internal region 
and the behavior of $\Delta \rho_p$ becomes closer to that of $^{44}$Ti.

In summary, the C-type density gives a more diffused surface than that
of the S-type in $^{44}$Ti because
the cluster configuration allows the occupation of the nodal $1p$ orbit
both for neutrons and protons.
For $^{52}$Ti, both the S- and C-types induce the enhancement
of the diffuseness because the S-type also fills in the $1p_{3/2}$ orbit
due to the Pauli principle from the $^{48}$Ca core.
The difference in the density profiles for the S- and C-type configurations
is found to be less drastic than in the case of $^{44}$Ti.
Investigation of the spectroscopic properties
  of these nuclei can corroborate this scenario,
  which can be achieved by using, e.g., the nucleon(s)
  knockout reactions~\cite{Gade08}.

\section{Conclusion}
\label{conclusion.sec}

We have studied the degree of the $\alpha$ clustering
in the ground state of $^{44}$Ti and $^{52}$Ti by
using fully antisymmetrized wave functions,
antisymmetrized quasi-cluster model (AQCM),
which can describe both the shell and cluster configurations
in a single scheme.
The characteristics of the density profiles are elucidated
by assuming the shell-model and $\alpha$-cluster-like configurations.
The nuclear surface is diffused by nodal single-particle orbits
induced by localized four-nucleons at the nuclear surface.
The difference between the shell and cluster configurations
becomes apparent for $^{44}$Ti,
while it is less for $^{52}$Ti because the shell model configuration
also has a diffused nuclear surface originating from the $1p_{3/2}$ orbit
due to the Pauli principle between the excess neutrons.

In this paper, we show two limits of shell and cluster configurations
and find that these two aspects can be distinguished by measuring
the proton-nucleus elastic differential cross section
up to the first peak position
as well as the nucleus-nucleus total reaction cross sections.
In reality, a nucleus consists of a mixture of these two limits.
Thus, these measurements will tell us dominant configurations
of the projectile nucleus, which offers a complementary tool to quantify
the existence of $\alpha$ cluster near the nuclear surface.

\acknowledgments
This work was in part supported by JSPS KAKENHI Grants
Nos.\ 18K03635, 22H01214, and 22K03618.
We acknowledge the Collaborative Research Program 2022, 
Information Initiative Center, Hokkaido University.

\end{document}